\documentclass[11pt,a4paper,nofootinbib,floatfix]{revtex4-1}
\usepackage[utf8]{inputenc}
\usepackage{amsmath}
\usepackage{centernot}
\usepackage{graphicx}
\usepackage{lipsum}
\usepackage{color}
\setcitestyle{numbers,square}

\begin{document}

\title{Shall we turn off the media? Global information can destroy local cooperation in the one-dimensional ring}

\author{Bilge Aydin\textsuperscript{1}}
\author{Marta Biondo\textsuperscript{2}}
\author{Deepak Gupta\textsuperscript{3}}
\author{Mattia Ivaldi\textsuperscript{2}}
\author{Francesca Lipari\textsuperscript{4,5}}
\author{Pablo Lozano\textsuperscript{5,6}}
\author{Francesco Parino\textsuperscript{7}}
\author{Ennio Bilancini\textsuperscript{8}}
\email{ennio.bilancini@imtlucca.it}

\author{Leonardo Boncinelli\textsuperscript{9}}
\author{Valerio Capraro\textsuperscript{10}}
\affiliation{}
\affiliation{\textsuperscript{1}Department of Urban Planning, Istanbul Technical University, Maslak, 34467 Sarıyer/İstanbul, Turkey}
\affiliation{\textsuperscript{2}Department of Physics, Università degli Studi di Torino, Via Pietro Giuria 1, 10125 Torino, Italy}
\affiliation{\textsuperscript{3}Dipartimento di Fisica G. Galilei, INFN, Università di Padova, Via Marzolo 8, 35131 Padova, Italy}
\affiliation{\textsuperscript{4}Department of Economics and Law, University of Rome, Via Marcantonio Colonna, 19, 00192 Roma, Italy}
\affiliation{\textsuperscript{5}Grupo Interdisciplinar de Sistemas Complejos (GISC)), Departamento de Matemáticas, Universidad Carlos III de Madrid, E-28911, Leganés, Madrid, Spain}
\affiliation{\textsuperscript{6}Unidad Mixta de Comportamiento y Complejidad Social UC3M-UV-UZ (UMICCS), Madrid, Spain}
\affiliation{\textsuperscript{7} ISI Foundation, Turin, Italy}
\affiliation{\textsuperscript{8}IMT School for Advanced Studies, Piazza San Francesco 19, 55100 Lucca, Italy}
\affiliation{\textsuperscript{9}Department of Economics and Management, University of Florence, Via delle Pandette, 32, 50127 Firenze, Italy}
\affiliation{\textsuperscript{10}Department of Economics, Middlesex University, The Burroughs, Hendon, London, UK}

\begin{center}
\fbox{
\begin{minipage}{\textwidth}
	\begin{center}
		\textbf{Important preliminary note:} This is a preliminary paper written in 72 hours as a challenge for the Complexity72h workshop (https://complexity72h.weebly.com). We apologize for the inaccuracies. An extended version will be posted as soon as possible.
	\end{center}
\end{minipage}
}
\end{center}

\begin{abstract}
\noindent In this paper we investigate the evolution of cooperation when the interaction structure is strictly local, and  hence fitness only depends on local behaviors, while the competition structure is partly global, and hence selection can happen also between distant agents. We explore this novel setup by means of numerical simulations in a model where agents are arranged in a one-dimensional ring. Preliminary results suggest that the extent of global comparison systematically favors defection under strong selection, while its effect under weak selection is less systematic, but overall still favors defection. Further, the extent of global comparison seems to reduce sensibly fixation times.
\end{abstract}

\maketitle

\section{Introduction}

\noindent Humans are unique in the animal kingdom for their capacity to cooperate in large groups with unrelated others. Bees, ants, and the mole naked rat all live in large societies, but individuals in the same society share a substantial degree of biological relatedness. Non-human primates live in groups made of unrelated individuals, but these groups are rather small. In contrast, humans live in societies made of thousands, and sometimes millions of individuals, who cooperate everyday building the schools, the hospitals, and all the institutions that make humans, on average, richer, healthier, and safer than ever before \cite{pinker2018enlightenment}. In fact, scholars from several disciplines have argued that the capacity to cooperate is what has made humans particularly successful as an animal species \cite{fehr2002altruistic,milinski2002reputation,gintis2003explaining,fehr2004social,nowak2006five,herrmann2008antisocial,capraro2013model,rand2013human,perc2017statistical}, and psychologists have even proposed that the psychological mechanism supporting cooperation, \emph{shared intentionality}, is what makes humans uniquely humans, as it is possessed by children, but not by great apes \cite{tomasello2005understanding}.

The evolution of cooperation among humans is however a puzzle, because cooperating requires individuals to pay a cost to benefit other individuals. Therefore, if evolution favors the fittest, non-cooperators should evolve at the price of cooperators. But this is not what we observe. What, then, makes humans so cooperative? Five fundamental rules have been proposed \cite{nowak2006five}, among which a prominent role has been played by \emph{network reciprocity}. Humans, indeed, do not usually interact at random, but they tend to interact with some individuals more often than others \cite{roca_cuesta09,Roca_2009,vilone2012social}. This may lead to the evolution of cooperation by reciprocity: I help you today because I expect you to help me tomorrow. This mechanism can create clusters of cooperators that can defend themselves from the invasion of defectors. Previous research using evolutionary game theory has indeed shown that network reciprocity can support the evolution of cooperation \cite{Axelrod81theevolution, Barker2017, milinski87, nowak_sigmund92, nowak_may92, maynardsmith82, hofbauer_sigmund98, gintis09}.

The work of Ohtsuki \textit{et al.} \cite{ohtsuki06}, in particular, has been receiving considerable attention. Here, the authors examine the evolution of cooperation in several networks and under several updating mechanisms. All the updating rules considered by Ohtsuki \textit{et al.} \cite{ohtsuki06} share one property: they are \emph{local}. After each time step, a random individual is chosen to update its strategy: this individual will maintain its own strategy or imitate one of its neighbors'. According to this model, therefore, both interactions and information are local: individuals interact with their neighbors and then update their strategy according to the information they acquire about their neighbors' strategy.

But modern societies are different. While it remains true that we mainly interact with our neighbors, we are constantly exposed to information regarding people we do not usually interact with. Social media, newspapers, television, and radio continuously inform us about the actions of people we have never seen before. We learn about extremely successful people as well as about extremely unsuccessful ones, and this information can influence the way we update our strategy in reality. In other words, in modern societies, information is not local, but has a global component. This raises an important question: Does global information favor or disfavor the emergence of cooperation, compared to local information, when interaction is still local?

Here we report a set of numerical simulations to study the evolution of cooperation in situations in which interactions are local, but information is, with some probability $p$, global. We let the parameter $p$ vary from $0$ (local information) to $1$ (fully global information). As a network structure, we consider the basic case of the one-dimensional ring, considered also by Ohtsuki \textit{et al.} \cite{ohtsuki06}. Following this work, nodes interact with their neighbors through a Prisoner's dilemma (PD). After every interaction, one node is selected at random to change its behavior (cooperate or defect), whereas another node is chosen to play the role of the source of information. With probability $p$ the source of information is random (global information); with probability $1-p$ the source of information is a neighbor of the selected node (local information). The selected node compares its payoff with the one of the source of information and then updates its behavior in agreement with the standard Fermi rule. Does $p$ favor or disfavor the evolution of cooperation? Does such an influence (if any) depend on the strength of selection?

To answer these questions, we compute the fixation probabilities of cooperation and defection using standard methods from evolutionary biology. We start from a Defective Society of size $N$, in which all nodes but one are initially defectors, and we compute the frequency in which the cooperator invades the society (fixation probability of cooperation); similarly, we start from a Cooperative Society in which all nodes but one are initially cooperators, and we compute the frequency in which the defector invades the society (fixation probability of defection). By comparing these probabilities with the the probability of fixation of the neutral character (which is $1/N$), we can first answer the question whether global information favor or disfavor cooperation and defection relative to the neutral character. Finally, comparing the fixation probability of cooperation with the fixation probability with defection, we can see which character among cooperation and defection is favored by selection.

\section{Methods}
\begin{figure}[ht]
\centering
\includegraphics[width=11cm]{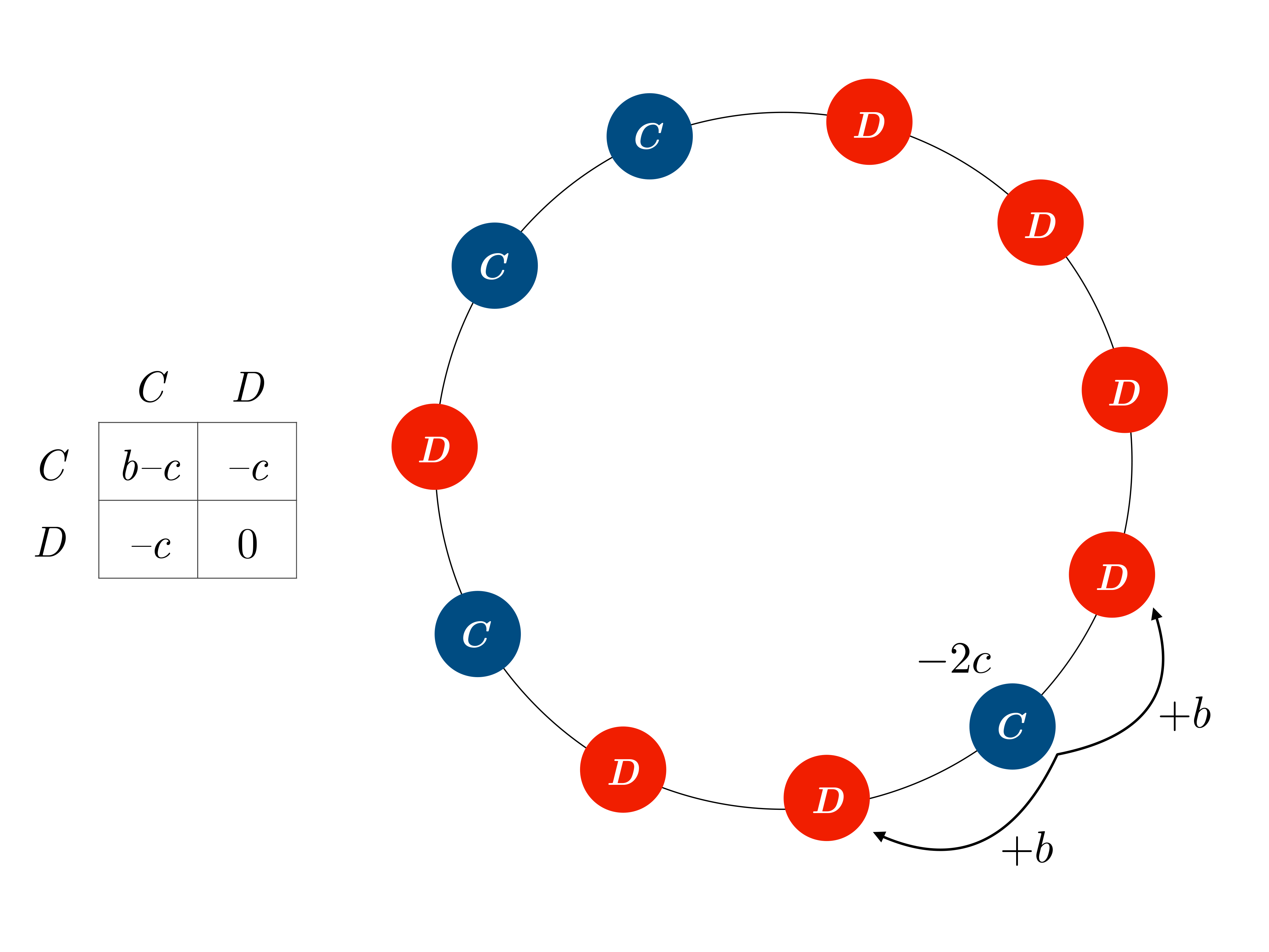}
\caption{The game. The matrix summarizes how payoffs are obtained: a cooperator pays $c$ for each neighbor, while both a defector and a cooperator get a benefit $b$ for each neighboring cooperator.}
\label{fig:dynamics}
\end{figure}

\noindent We consider the Prisoner's Dilemma game among $N$ players, who interact with their neighbors in a one-dimensional ring. In the PD, a player can either cooperate (C) or defect (D). Cooperating means paying a cost $c$ to give a benefit $b > c$ to the other player 
(see Figure \ref{fig:dynamics}). 


\begin{figure}[h]
\centering
\includegraphics[width = 6cm]{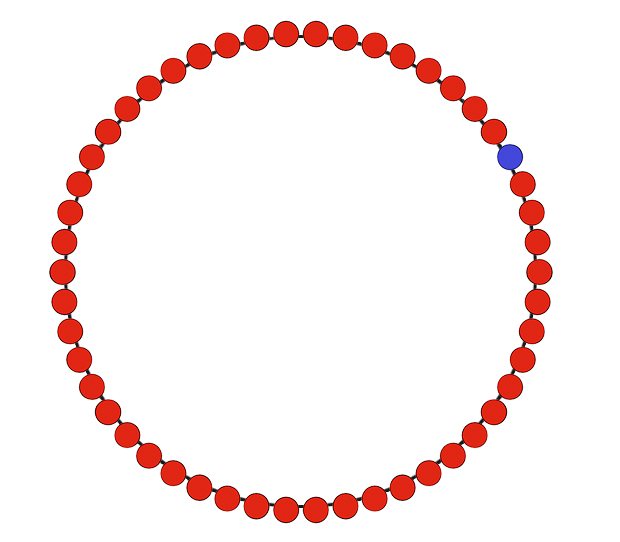}
\includegraphics[width = 6cm]{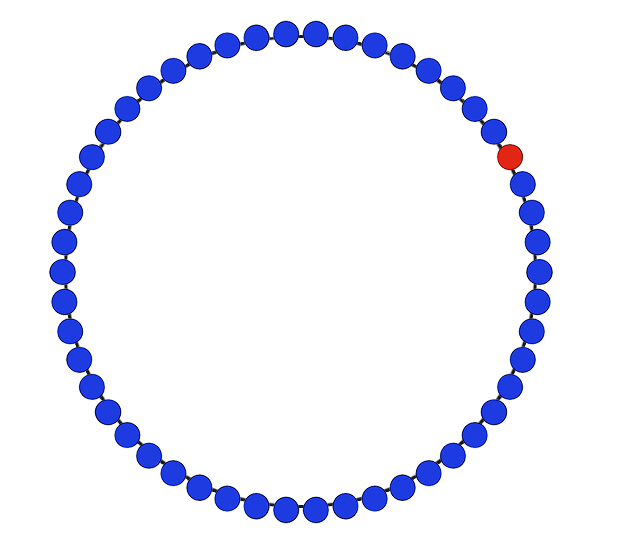}
\caption{The initializing populations. The red ring on the left represent the Defecting-Society where one single cooperators is surrounded by defectors. The blue ring, instead, represent the Cooperative- Society.}
\label{fig:network}
\end{figure}

We initialize two populations:  the Defecting-Society in which all agents but (blue) one are defectors and the Cooperating-Society in which all agents are cooperators except for one (red) as showed in Figure \ref{fig:network}. 

To compute the probability of fixation of cooperators (defectors), for a given realization, we initialize the system by labelling $N-1$ nodes as defectors (cooperators) and the remaining node as cooperator (defectors). The simulations are run with discrete time ($t=0, 1, 2, ...$). At time $t=1$, we do the following steps. First, we select a node $j$ with probability $1/N$. Then, with probability $p$, a random node $l\neq j$ is chosen to play the role of the source of information (global information); with probability $1-p$, the nearest neighbours ($j-1$ and $j+1$) of $j$ are selected to play the role of sources of information (local information). If the selected nodes have all the same strategy of the original node, then we end the step and start over by selecting another random node $j$. Otherwise, we compute the average payoff of the defectors, $\langle \pi_D\rangle$, and that of the cooperators, $\langle \pi_C\rangle$, for all the selected nodes. Then, if node $j$ was a cooperator, then it updates its strategy from $C$ to $D$ with probability
\begin{align}
\mathcal{P}\{C \longrightarrow D\}=\dfrac{1}{1+e^{w(\langle \pi_C\rangle-\langle \pi_D\rangle) }};
\label{fate}
\end{align}
else if $j$ was a defector, then $j$ changes its strategy from $D$ to $C$ with probability $\mathcal{Q}\{D \longrightarrow C\}=1-\mathcal{P}\{C \longrightarrow D\}$. ($w$ represents the strength of selection: if $w\to0$, payoffs cease to matter and strategies change at random; if $w\to\infty$, players change strategy only if they acquire information from a more successful node.) 

\begin{figure} [h]
    \centering
    \includegraphics[width=15cm]{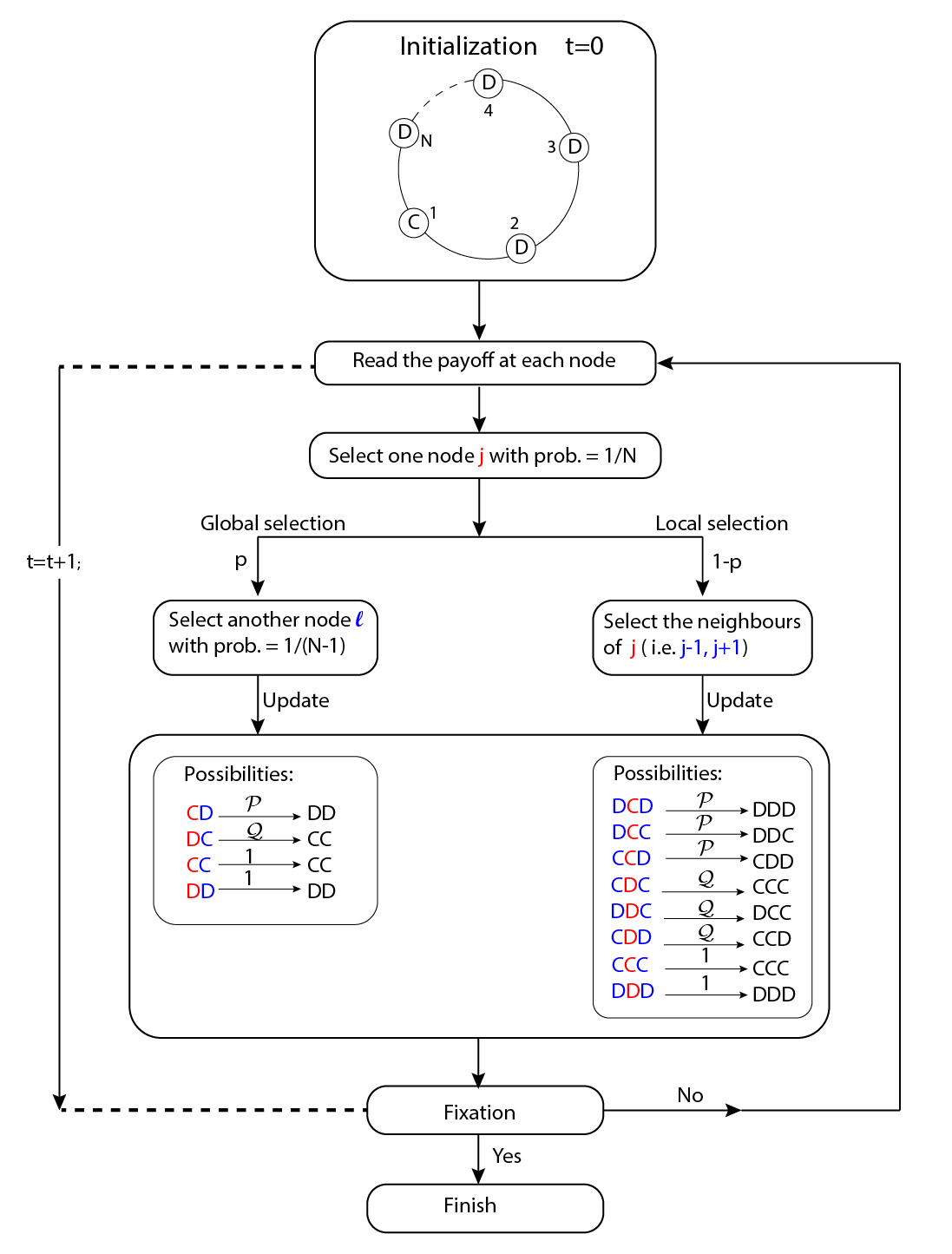}
    \caption{Flow chart for the numerical simulation to compute the fixation probability of cooperation. Red color letters indicate the first selected nodes, whereas blue letters are those chosen according to the global or local information. The probability $\mathcal{P}\{C \longrightarrow D\}$ and $\mathcal{Q}\{D \longrightarrow C\}$ to accept the change are computed as in Eq.~\eqref{fate}.}
    \label{fig:FC}
\end{figure}

Similarly, in the case of the global dynamics, the probability that the node $j$ updates its strategy is evaluated using the same rule (Eq.~\eqref{fate}), but this time the comparison is done with $l$, instead of $j-1$ and $j+1$. Finally, the time is incremented by one unit. We recursively perform the same realization until the fixation for the cooperation (defection) is achieved, that is, all the nodes become cooperators (defectors). 
The scheme of the numerical simulation to compute the fixation of the cooperation is shown in Fig.~\ref{fig:FC}.

\section{Results}




\noindent By means of simulations we explore if natural selection can favor cooperation on a one-dimensional ring for different level of local and global information. In order to do this, we need to calculate the probability that a single cooperator starting in a random position turns the whole population from defection to cooperation in the Defecting Society. We also calculate the reciprocal fixation probability of a single defector in a population of cooperators turning the whole population to defection, and compare the two fixation probabilities against each other and against the general case of well-mixed population. Furthermore we calculate also the convergence time that the society needs to reach its final configuration of all cooperators (all defectors) with respect to the probability of going local.\footnote{Since the value of $w=1$ is a critical one, we run different number of simulations, $\mathcal{R}=5000$ for fixation probability and $\mathcal{R}=10000$ for the convergence time.}

\begin{figure}[ht]
    \centering
    \includegraphics[width=8cm]{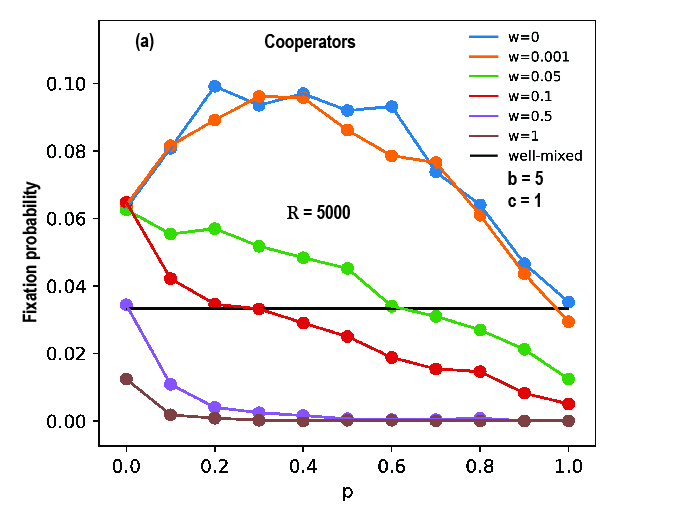}
    \includegraphics[width=8cm]{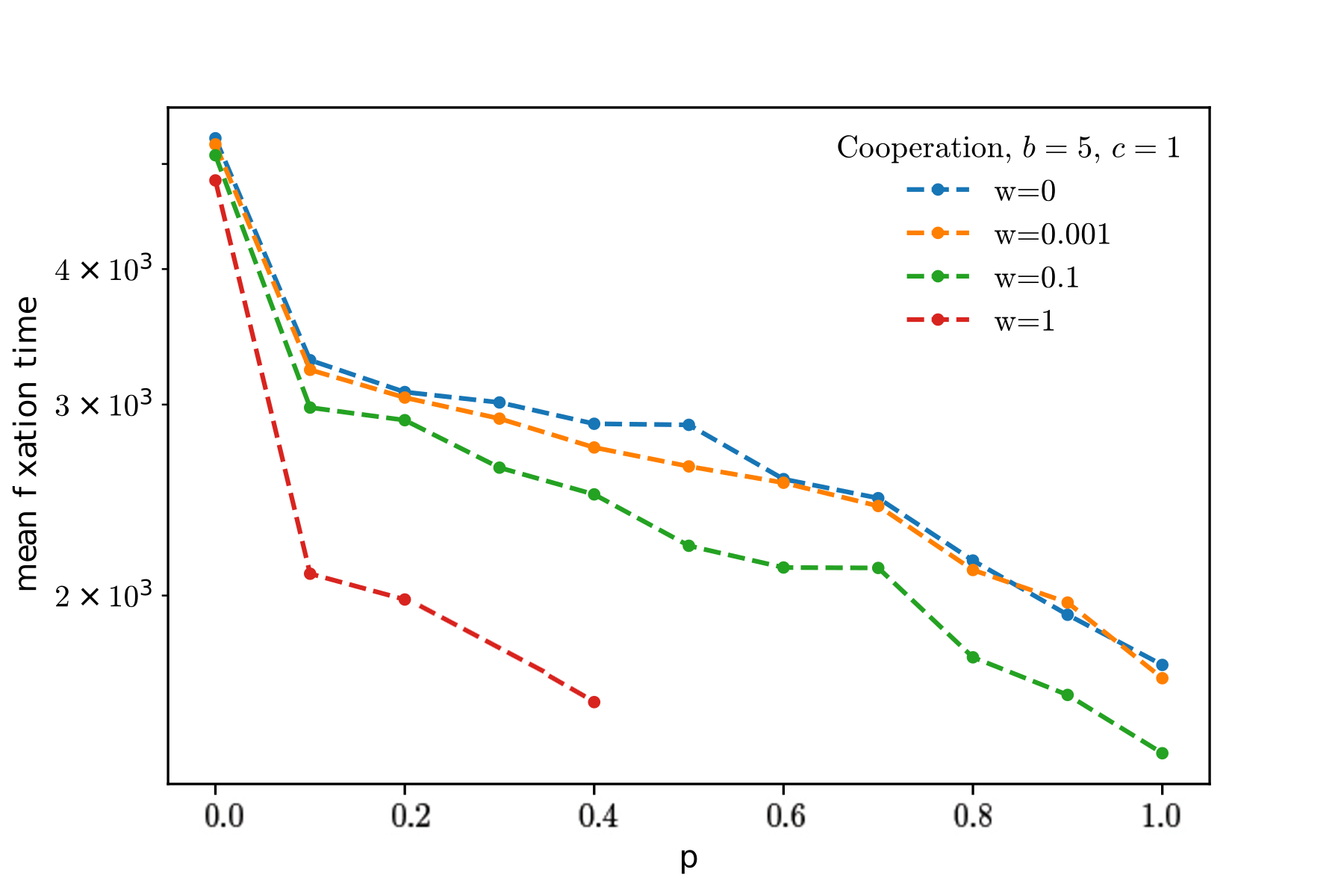}
    \caption{\textbf{Evolution of Cooperation.} Results for the fixation probability of cooperation in Defecting Society. The simulations were run for $N=30$ and $b=5$. Panel (a) on the left shows the evolution of the fixation probability of cooperators for different values of $p$ and of the strength of selection $w$, compared to the results obtained in the well-mixed population case. Panel (b) on the right shows average convergence time for different values of $p$ and of the strength of selection $w$.}
    \label{fig:fix_coop}
\end{figure}

\textbf{Defecting Society}:
We first present the results of the analysis of the Defecting Society in which all nodes but one are defectors. How likely is it that this society will reach a configuration in which all the agents are cooperators? How does this likelihood vary in function of the probability of receiving global information? We study this question for several selection strengths. Figure \ref{fig:fix_coop} panel (a) shows the fixation probability of cooperation as a function of acquiring global information, and for distinct level of selection, from weak to strong. We compare the resulting trends to the fixation probability of a neutral character (represented in the figure with gray straight line). It is visually evident that the effect of global information on the fixation probability of cooperation is non trivial, as it turns out to be sometimes monotone, and sometimes not. In particular, when the strength of selection is strong ($w=0.5, w=1$), the fixation probability of cooperation quickly goes to zero as $p$ increases. Therefore, with strong selection, global information has a detrimental effect on the fixation probability of cooperation. The same detrimental effect is observed also for mild levels of selection: when $w=0.1$ and $w=0.05$, the probability of fixation of cooperation still decreases with $p$, but it remains relatively high, and anyway higher than the fixation probability of a neutral character until relatively high values of $p$, especially for $w=0.05$. The effect of acquiring global information becomes non-trivial, and evidently non-linear when weak selection is at play ($w=0.01$) and in the limit case $(w=0)$, in which nodes update their strategy at random. In this case, the probability of fixation of cooperation first increases for relatively small values of $p$, then, after reaching a maximum around $p=0.3$, it starts decreasing, giving rise to an overall inverted-U effect. Since the initial (for $p=0$) fixation probability of cooperation is already greater than the fixation probability of the neutral character (which is $\frac1N\sim0.03$), a consequence of this inverted-U effect is that the probability of fixation of cooperation remains greater than $1/30$ for very large values of $p$, approaching $1$. A potential interpretation for these results is that when $p$ approaches $1$, the graph becomes conceptually similar to a complete graph, in which, although nodes still play only with their neighbors, they acquire information about the full graph. In the full graph, which corresponds to the well-mixed population, defection wins, and this might explain why, for large $p$ we observe a decay on the fixation probability of cooperation, independently of the strength of selection.


Next, we analyze the time needed for the system to reach the final configuration, as function of the local (or global) interaction. Figure \ref{fig:fix_coop} panel (b) displays the mean fixation time of cooperation as a function of $p$. In case of weak and mild selection, the average convergence time decreases with the probability of receiving global information. When selection is strong ($w=1$) and $p>0.4$, it is almost impossible to reach total cooperation, so the mean fixation time is not defined.


\begin{figure}[ht]
    \centering
    \includegraphics[width=8cm]{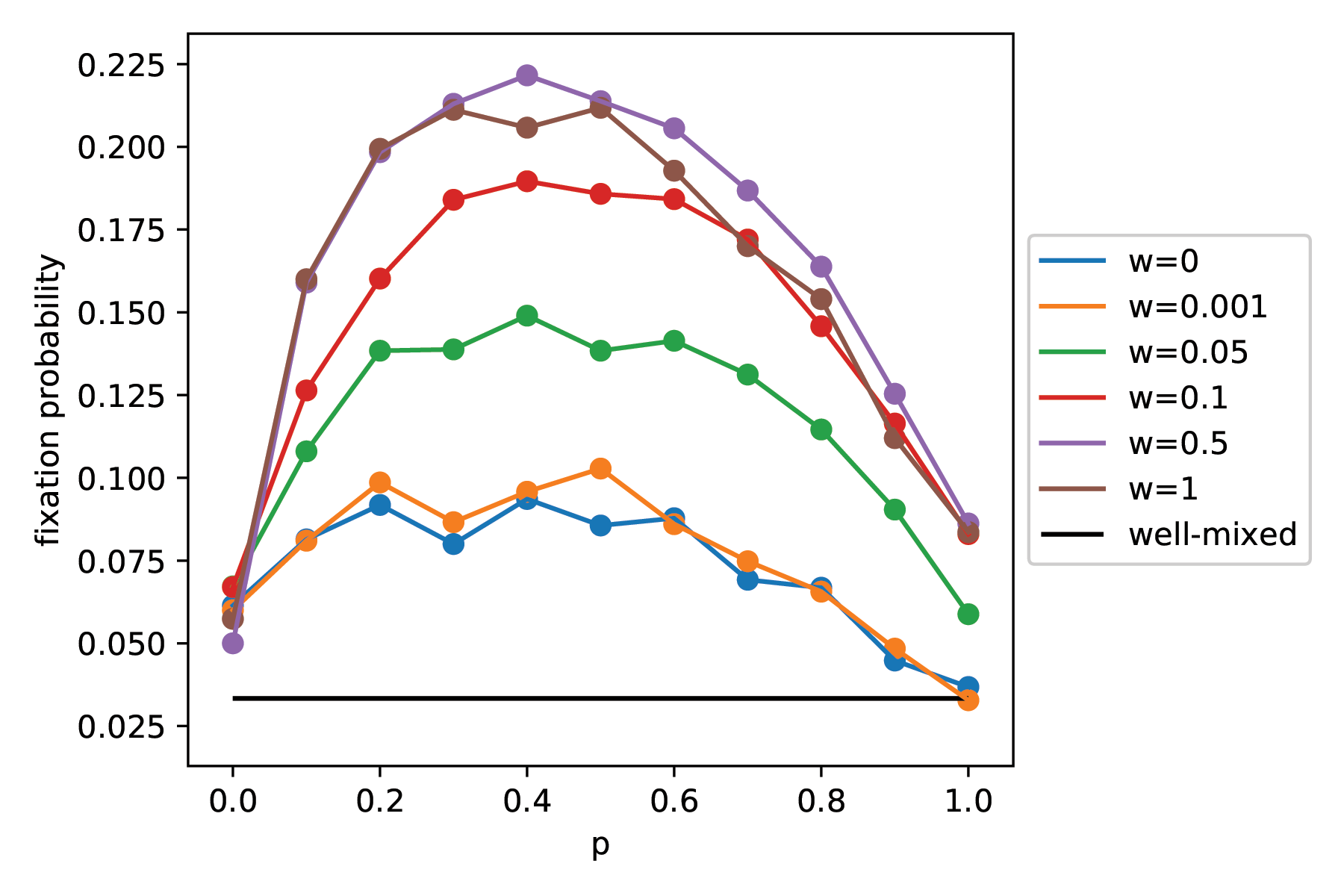}
    \includegraphics[width=8cm]{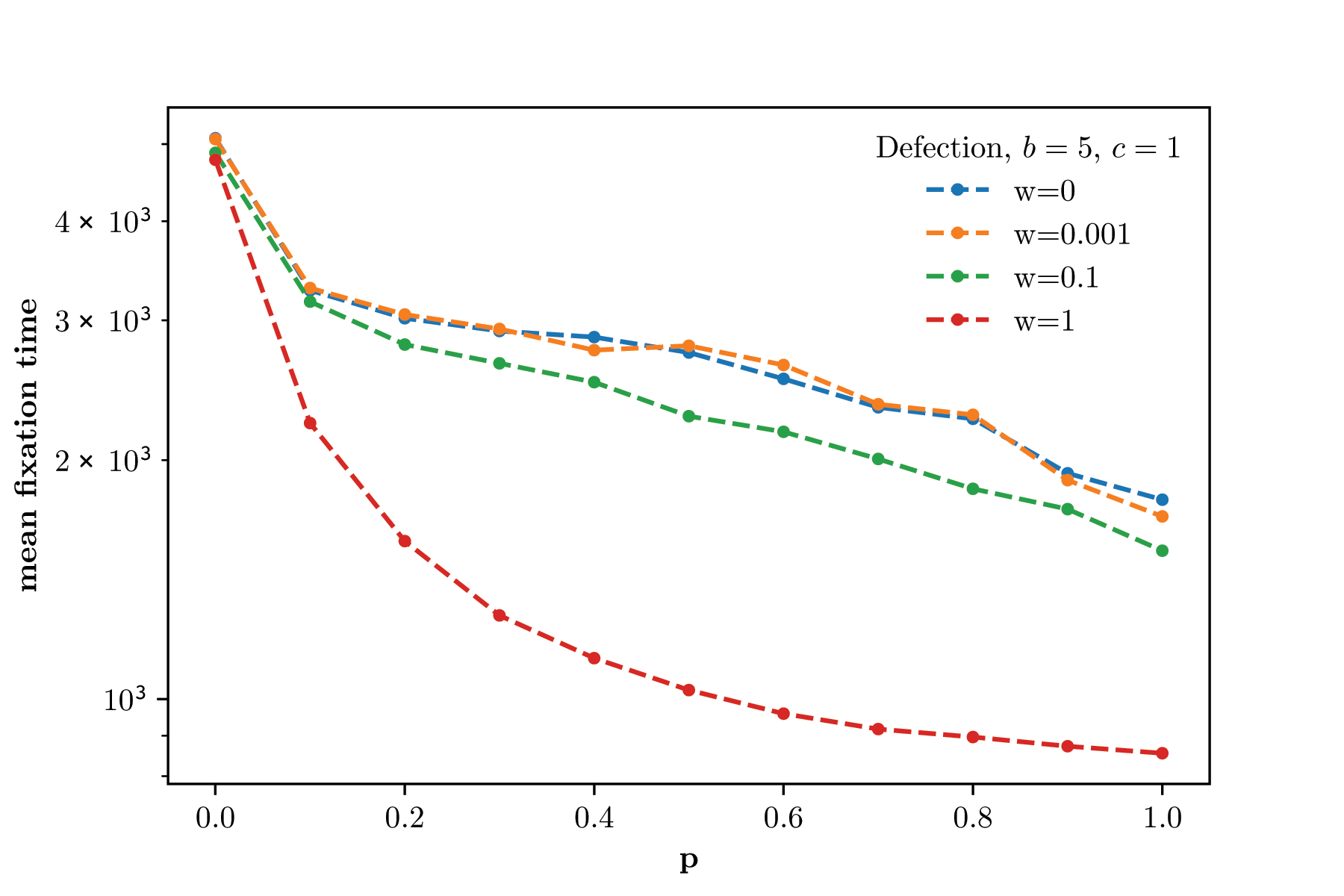}
    \caption{\textbf{Evolution of Defection.} Results for the fixation probability of defection in Cooperative Society. The simulations were run for $N=30$ and $b=5$. Panel (a) on the left illustrates the evolution of the average time needed to reach total defection for different values of $p$ and of the strength of selection $w$. Panel (b) on the right shows average convergence time for different values of $p$ and of the strength of selection $w$.}
    \label{fig:fix_def}
\end{figure}

\textbf{Cooperative Society}: 
Figure \ref{fig:fix_def} shows the results of the simulation for the Cooperative Society, in which all the nodes but one are initially cooperators. How likely is it that this society will reach a configuration in which all the agents are defectors? How does this likelihood vary in function of the probability of receiving global information? The left panel of Figure \ref{fig:fix_def} shows the trend of the fixation probability of defection as a function of the probability of acquiring global information, and for distinct level of selection. Also here, we compare the resulting trends to the fixation probability of a neutral character (i.e. gray straight line). In contrast to the case of Defecting Society, the fixation probability of defectors are have an inverted-U shape, independently of $p$, and they remain constantly higher than the fixation probability of the neutral character. More specifically, when $p=0$, in which the agents access only local information, the differences among the fixation probabilities are minimal. From $p>0$ the six curves start to diverge. When weak selection is at work, since the payoffs have little impact on evolution, the fixation probability of defectors is restrained even in case of global information, that is, increasing the probability of receiving global information favors the spread of defection, but at a small rate. The opposite occurs at mild and strong levels of selections. In this case, the effect of the payoffs on selection is greater, and this results in the fact that the probability of reaching a defecting state increases. Finally, as $p$ approaches 1, increases to the point that information is completely random therefore the defectors become more likely to acquire information of successful cooperators (because the initial state of the society is cooperative), and this leads to a decrease in the fixation probability of defectors. 

Figure \ref{fig:fix_def} panel (b) displays the mean fixation time of the defectors as function of $p$. In case of weak and mild selection the average convergence time decreases with $p$ with an almost linear trend. When the selection is strong ($w=1$) fixation time decreases hyperbolically.

\begin{figure}[ht]
    \centering
    \includegraphics[width=8cm]{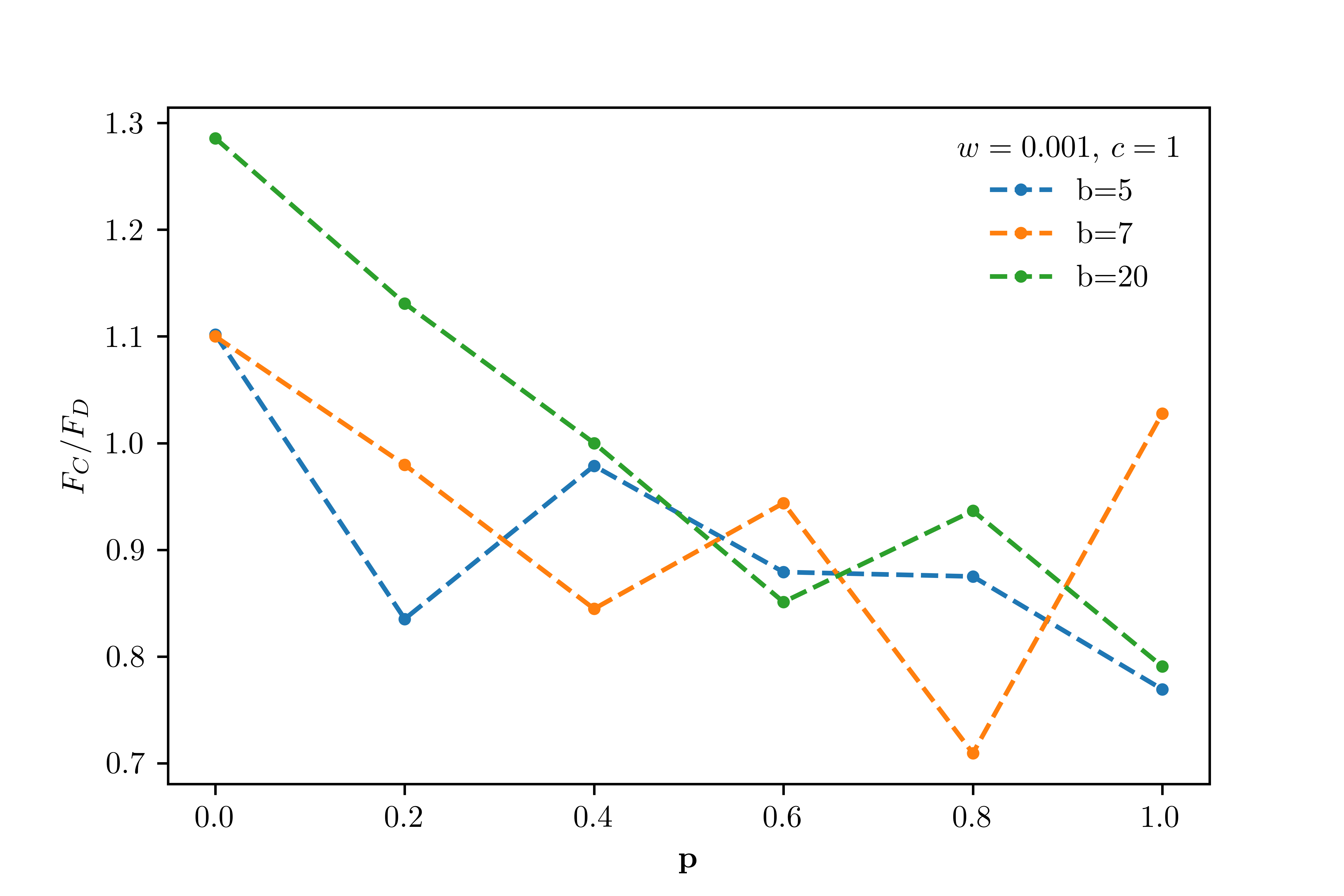}
    \includegraphics[width=8cm]{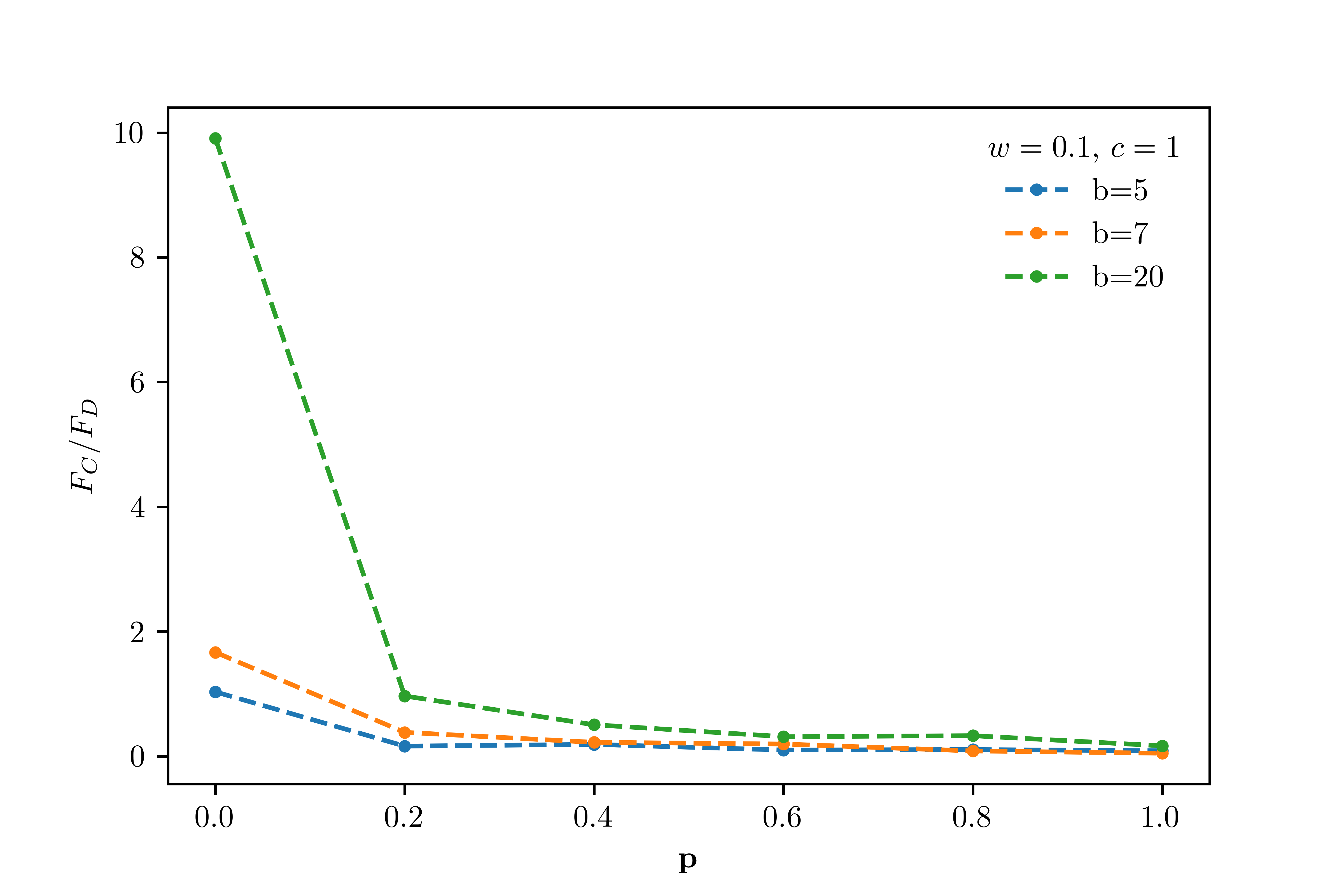}
    \caption{\textbf{Comparison between fixation probabilities.} Results for the ratio of fixation probabilities as a function of the probability of receiving global information, for different values of $b$. The simulations were run for $N=30$ and $b=5$. Panel (a) refers to a case of weak selection, while in Panel (b) reports a case of mild selection. In this case the number of simulations is $\mathcal{R}=1000$.}
    \label{fig:fixation_comparison}
\end{figure}

\textbf{Relative fixation}:
Here we compare the probability of fixation of cooperation ($F_C$) with the probability of fixation of defection ($F_D$), as a function of $p$ and for weak and mild selection, separately (the case of strong selection is obvious and we do not report it in the paper), to see whether global information favors the evolution of cooperation relative to the evolution of defection. Specifically, if $F_C/F_D>1$, then the evolution of cooperation is favored over the evolution of defection; else if, $F_C/F_D<1$, then the evolution of defection is favored over the evolution of cooperation. Figure \ref{fig:fixation_comparison} reports the relative fixation probability of cooperation or defection in the case of weak selection (left panel, $w=0.001$) and mild selection (right panel, $w=0.1$). We conducted simulation for several values of the benefit for cooperation $b$. The results are somewhat interesting, especially in the case of weak selection. In case of mild selection (and more so in case of strong selection), the ratio $F_C/F_D$ decreases, thus providing evidence that global information is detrimental for the evolution of cooperation relative to the evolution of defection. In case of weak selection, however, the results are more interesting. Indeed, on the one hand, also in case of weak selection, global information is detrimental to the evolution of cooperation in the sense that $F_C/F_D$ for $p>0$ is always smaller than its value for $p=0$. However, this time $F_C/F_D$ does not decrease with $p$, but follows a non-linear trend. For example, for $b=5$ and $p=0.1$, we note a quick decrease, respect to $p=0$, of $F_C/F_D$. However, already for $p=0.2$, the value of $F_C/F_D$ increases back to values close to 1. A similar alternating behavior characterizes the other values of $p$ and the other values of $b$. 

\section{Discussion}
\noindent Here we studied whether keeping interaction local and allowing comparison to go global has an impact on the evolution of cooperation and defection in a population of agents playing a prisoner's dilemma game. As a spatial structure, we considered the one-dimensional ring, the basic network considered also in the seminal work by Ohtsuki \emph{et al.}~\cite{ohtsuki06}. Numerical simulations with varying strength of selection and varying probability of acquiring global information reveal that, overall, global information is detrimental for cooperation, especially under strong selection. 

The details of the evolution are however complex and non-trivial. While, under mild selection, global information destroys the evolution of cooperation very quickly and monotonically, resulting in a decreasing ratio between the fixation probability of cooperation and that of defection, $F_C/F_D$, under weak selection the ratio $F_C/F_D$ varies with $p$ following a non-linear trend. However, for the value of $F_C/F_D$ for $p>0$ is always smaller than its value for $p=0$, suggesting that global information is detrimental for local cooperation also in the case of weak selection.

On the ground of intuition, the overall effect of global information on the likelihood that cooperation gets favored by evolution appears uncertain. Indeed, global information favors cooperation over defection in that the comparison between a cooperator and a defector does not involve, in general, neighbors, which means that the defector does not benefit from the cooperator. At the same time, however, global information does allow for the rising of cooperators in non-clusterized structures, which reduces assortativity and, by doing so, also the relative advantage of cooperators over defectors. Our results seem to suggest that the second effect prevails, determining an overall negative impact of global information on the evolutionary success of cooperation. We stress, however, that our simulations are confined to the case of the one-dimensional ring, and the effects of global information need to be explored in more general interaction structures. For instance, one should explore greater dimensional lattices and also non regular graphs. Further, we modeled the flow of global information in a very extreme and ad hoc way. A reasonable alternative could be to have global information coming from neighbors of neighbors instead of a randomly selected distant agent. This could actually favor cooperation by reducing the likelihood that a cooperator is found outside a cluster of cooperators.


The fact that global information reduces monotonically fixation times may be due to the larger set of agents that can potentially change behavior in the presence of global information as opposed to local information. Indeed, when information is local, only agents at the boundaries between the cluster of cooperators and the cluster of defectors can actually switch action, while others are prevented simply because they do not observe a behavior different from their own. When information is global, instead, any agent has a positive probability to observe a behavior different from its own at every state which is not monomorphic.

\hfill \break
\textbf{Acknowledgments:} This work is the output of the Complexity72h workshop, held at IMT School in Lucca, 17-21 June 2019. https://complexity72h.weebly.com/

\bibliography{./jobname}

\end{document}